# The impact of analytical outage modeling on expansion planning problems in the area of power systems

Stamatis Tsianikas[a], Nooshin Yousefi[a], Jian Zhou[a], and David W. Coit[a]

[a] Department of Industrial and Systems Engineering, Rutgers University, Piscataway, NJ 08854-8065, USA

## Abstract

Expansion planning problems refer to the monetary and unit investment needed for energy production or storage. An inherent element in these problems is the element of stochasticity in various aspects, such as the generation output of the units, climate change or frequency and duration of grid outages. Especially for the latter one, outage modeling is crucial to be carefully considered when designing systems with distributed generation at their core, such as microgrids. An element that it is observed to be missing in the literature studies so far, is the existence of accurate probabilistic modeling for outage events. In most studies so far, a single statistical distribution is used, such as a Poisson Process. However, by taking a closer look at the real outage data provided by the state of NY (Service, 2017), it is observed that the outages do not seem to come from the same distribution. In some years, there is a huge spike in the average duration per outage and this is because of catastrophic events. Therefore, in this study we propose and test an alternative modeling for outage events. This alternative scheme will be based on the premise that outages can be broadly classified into two categories: regular and severe. Under this taxonomy, it can still be assumed that each type of events follows a Poisson Process but outages, in general, follow a Poisson Process which is truly a superposition of these two types. For a better illustration of the abovementioned modifications, Fig. 1 shows how the distribution of the outage duration can potentially change. A reinforcement learning approach is used to solve the expansion planning problem and real location-specific data are used. The results verify our initial hypothesis and show that the optimization results are significantly affected by the outage modeling. Moreover, another key finding of this work is that the mix of storage technologies also changes with a different outage modeling approach. To sum up, modeling accurately the grid outage events and measuring directly the reliability performance of an energy system during catastrophic failures could provide invaluable tools and insights that could therefore be used for the best possible preparation for this type of outages.

## 1. Introduction

The functioning of modern society, including personal life and industry, is increasingly dependent on continuous power supply. Thus, energy systems installed with distributed energy generation (DER) are becoming more and more popular since they are able to meet stochastic energy demand and have other advantages ranging from power resilience to renewable integration [1]. Microgrid is one the representative applications of this kind. Given that many energy storage options are more and more economical, determining which storage technologies to invest in, as well as the appropriate timing and capacity becomes critical [2, 3]. It leads to the optimization of expansion planning problem. This problem involves a lot of stochasticity regarding diverse practical aspects, such as uncertain nature of renewable energy resources and grid outages, which has been proved in exerting great impacts on expansion planning optimization and on the optimal design of energy storage systems [4]. Because of the increasing complexity and scale of power grids, cascading failures has become a common failure phenomenon in power grids [5]. It not only aggravates the influence of grid outages [6], but also increase the stochasticity of grid outages. Efforts have been made to mitigate the loss caused by cascading failure in power grids [7-9] while it still occurs.

As a result, grid outage modeling is apparently of great significance when planning new energy systems with DER. Modeling grid outages accurately can improve the effectiveness of these energy systems on reducing grid outage damage and enhancing energy resilience. Many works are conducted on grid outage modeling, for instance, many models are developed according to complex network theory to describe failure propagation behavior in power grids, which usually result in catastrophic events [10-13]. Single statistical distribution, like Poisson Process, is also applied to model grid outages in many studies. However, by taking a closer look at real data of grid outages, it is observed that grid outages do not seem to follow the same distribution. As noted, grid outages resulting from cascading failures can cause devastating loss and a huge spike in the average duration per outage. Thus, we proposed a new modeling approach for grid outages. For validation, a reinforcement learning approach is used to solve the problem for both current and proposed outage models and compare the results.

## 2. Problem formulation

The purpose of this work is to highlight the significance of accurate outage modeling when solving optimization problems in the area of expansion planning of energy systems. In order to do so, it is needed to compare two different modeling approaches for the outage events in a microgrid and explore how the optimization results are affected. The problem formulation and the mathematical model are adopted from [14] and [1]. The objective of this problem is to derive the optimal storage expansion plans for a specific microgrid in a predetermined time horizon. By specific microgrid, it is meant that the location, power plants and technological equipment of the microgrid are given and not allowed to change. However, the decision maker can choose between a number of storage options and a number of different capacity levels. Therefore, the problem melts down to finding the optimal strategy on when, how much and which technology to use when expanding the storage capability of a microgrid.

### 2.1 Nomenclature

| | |
|---|---|
| $S$ | State set |
| $A$ | Action set |
| $f$ | Transition function |
| $R$ | Reward function |
| $S^{tf}$ | Timing feature of the state set |
| $S^{ef}$ | External feature of the state set |
| $S^{if}$ | Internal feature of the state set |
| $K$ | Number of decision periods |
| $SU$ | Set for available storage units |
| $SC$ | Set for characteristics of storage units |
| $SL$ | Set of available storage capacity levels |
| $N(t)$ | Single Poisson Process |
| $N_1(t)$ | Poisson Process for the regular outage events |
| $N_2(t)$ | Poisson Process for the severe outage events |
| $N'(t)$ | Superposed Poisson Process |
| $Z_n$ | Type of the $n^{th}$ grid outage event |
| $T$ | Grid outage duration, hours |

### 2.2 MDPs and Q-learning for deriving the optimal strategy

In order to derive this optimal strategy, the notion of Markov Decision Processes (MDPs) is used. The critical elements in an MDP are the agent and the environment. The agent is basically the decision-maker in the problem and the one who is responsible for learning. The environment is all the things that the agent should interact with, in order to get information [15]. In MDPs, there is always a tuple *(S,A,f,R)* which can fully define it. *S*, *A* refer to the state and action sets of the problem and *f*, *R* correspond to the transition and reward function. To be more specific, the state set of this problem is defined as a collection of three features: the timing feature, the external feature and the internal feature. The timing feature corresponds to the decision period, the external feature to all the various storage characteristics that the decision maker cannot influence (i.e. price, efficiency etc.) and finally the internal feature to the storage capacity installed in the system for a specific technology. This can be seen in Equation 1:

$$S = S^{tf} \times S^{ef} \times S^{if} \tag{1}$$

where: $\quad s^{tf} \in S^{tf} = \{1, 2, ..., K\}$

$\vec{s}^{ef} = \left(s^{ef}_{i,j}, \forall i \in SU, j \in SC\right) \in S^{ef}$

$\vec{s}^{if} = \left(s^{if}_{i}, \forall i \in SU\right) \in S^{if}, \ \forall i \in SU, s^{tf} \in S^{tf}$

where $S^{tf}$, $S^{ef}$ and $S^{if}$ correspond to the timing, external and internal features described above. *K* is the number of decision periods in the problem, *SU* is the set of the available storage units and *SC* is the set of storage characteristics for each storage unit.

Furthermore, the action set of the problem is clearly defined based on the actions that the agent is allowed to take. In every decision period, it is possible to do nothing or install a specific storage unit in the desired level. The level should come from a fixed set of available storage levels. The reason for this comes from the discrete state and time assumptions of the MDP framework. The action set is defined in Equation 2:

$$\vec{a} = (a_{i,l}, \forall i \in SU, l \in SL) \in A \tag{2}$$

s.t. $a_{i,l} \in \{0,1\}, \forall i \in SU, j \in SL$

$\sum_{i \in SU, l \in SL} a_{i,l} \in \{0,1\}$

where *SL* is the set of available storage levels. It should be noted here that the constraint refers to the fact that the agent is not allowed to take more than one actions per decision period.

Concerning the transition function *f* of the problem, this should be defined in accordance with the state set definition given in Equation 1. Of course, it is necessary to preserve the Markovian property for every component of the transition function. Therefore, the external feature component of the transition function is actually a collection of Discrete Time Markov Chains (DTMCs), one for each characteristic of every available storage unit. The other two parts, are much simpler to derive. The timing component is an incremental by-one operation and the transition of the internal component depends on the actions taken by the agent.

Finally, the reward function *R* is crucial since it affects the way that the environment sends back signals to the agent, i.e. the entity that makes the decisions. Using these signals, the agent is able to be guided towards better policies and finally converge to the optimal one. In this problem, the reward function consists of two components: the investment cost and the outage cost. It is clear that there is a tradeoff here: the agent should take actions to install large enough storage capacity to protect against a significant number of outages, however not large enough that could result in a large portion of it staying idle and unused. The modeling approach that is chosen for the outage events is very crucial in calculating the cost component and affects significantly the optimization results.

The algorithm used in this work for solving this sequential decision making problem is a variant of probably one of the most famous reinforcement learning algorithms, initially proposed in [16] and called Q-learning. Q-learning is a model-free, off-policy learning algorithm that depends on the well-known Bellman equation. After the training of the algorithm is finished and convergence has been achieved, the result of Q-learning is a so-called Q-table. Q-table has all the problem states in rows and all the available actions in columns. The elements in the table are called Q-values and are used to determine which action is the optimal one for each state. The main difference between the classical Q-learning approach and the variant used in this work is the existence of a metamodel for calculating the outage cost component of the reward function.

### 2.3 Current and proposed probabilistic modeling approach for outage events

Firstly, the probabilistic modeling approach used in [1] and [17] should be defined. Let $\{N(t), t \in [0, +\infty)\}$ be the counting process which defines the outage events in the system. $N(t)$ is considered to be a Poisson Process with rate $\lambda$ and therefore the number of outages at any given time $\tau > 0$ follows a Poisson distribution with rate $\lambda \tau$. Moreover, the duration of each outage $T$ is again following a shifted Poisson distribution with a rate $\kappa$. It becomes clear from the definition that all outages are assumed to be independent and identically distributed. While the assumption about the independence of each outage may seem reasonable in some cases, specific attention should be paid to the distribution under which various outages are happening. In order to do that, someone should check the data for the average duration per interruption (CAIDI) that the NY state provided in Table 1, for the case of PSEG-LI [18]:

Table 1: CAIDI data provided by NY state for PSEG-LI, years 2012-2017.

| Year<br>Duration | 2012 | 2013 | 2014 | 2015 | 2016 | 2017 |
|---|---|---|---|---|---|---|

| CAIDI (hrs/int) | 22.55 | 1.65 | 1.42 | 1.95 | 1.46 | 1.70 |

By examining the data provided in Table 1 it is observed that the outages do not seem to come from the same distribution. There is a huge spike in the average duration per outage in the year 2012 and the most obvious reason for that is the devastating Hurricane Sandy that happened on October 22, 2012, and affected a vast majority of the US Northeast area for prolonged periods. Therefore, it may seem reasonable to propose and test an alternative modeling for outage events. This alternative scheme will be based on the premise that outages can be broadly classified into two categories: regular and severe.

Under this taxonomy, it can still be assumed that each type of events follows a Poisson Process and outages, in general, follow a Poisson Process which is truly a superposition of these two types. Therefore, if $\{N_1(t), t \in [0, +\infty)\}$ with rate $\lambda_1$ is a counting (Poisson) process for the regular outage events and $\{N_2(t), t \in [0, +\infty)\}$ with rate $\lambda_2$ is a counting (Poisson) process for the severe outage events, $N'(t) = N_1(t) + N_2(t)$ is a superposed Poisson Process with rate $\lambda = \lambda_1 + \lambda_2$. In that case, the probability that a random outage event comes from either of these processes should also be defined. Therefore, let $\Pr\{Z_n = i\} = \frac{\lambda_i}{\lambda}$ where $Z_n$ is the type of $n^{th}$ event and of course $i \in \{1, 2\}$ in this case. Of course, the duration of the outages can still be assumed to follow a shifted Poisson distribution but in a similar way, two distinct random variables are now defined, such as $T_1, T_2$ with respective rates $\kappa_1, \kappa_2$. For a better illustration of the abovementioned modifications, Fig. 1 shows how the distribution of the outage duration can potentially change:

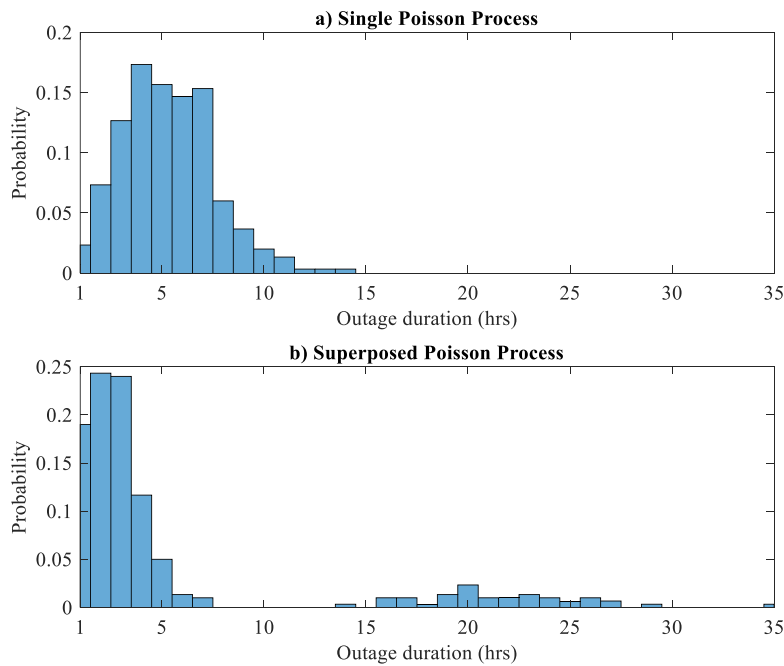

Fig. 1. Distribution of outage duration using two different probabilistic modeling approaches

By observing Fig. 1, it is more than clear that the distribution of the outage duration is significantly altered, even though the mean duration may have stayed the same.

## 3. Case study

In this section, a case study is presented for better illustration of the aforementioned procedures. A microgrid in Westhampton, NY is selected for the demonstration purposes and is subject to grid outages. This microgrid consists of several facilities and actual demand [19] and solar irradiance data [20] are used. The storage options existing in *SU* are four: Li-ion battery, lead-acid battery, vanadium redox battery, and flywheel storage system. The agent is able to install any of these storage units at any of three predetermined levels in *SL*. Finally, the stochastic component considered in $S^{ef}$ is the price of the storage units, which is modeled by a DTMC, different for each storage unit.

### 3.1 Numerical assumptions

Regarding the numerical assumptions of the case study, there are 2 hospitals, 5 schools and 300 residential buildings in the microgrid. Each facility type has its own criticality and energy consumption. The predetermined levels in *SL* are: 300, 1000 and 3000 kWh. The DTMC for the price of each storage unit is defined to have four states, a probability of 0.7 for the transition from state $s$ to $s'$ and a probability of 0.3 for staying in the same state. Concerning the Q-learning algorithm, there is a total of $10^6$ training episodes for each approach and also linearly decaying learning and exploration/exploitation rates are used. For more information on the exact numerical assumptions for the power plants, the storage characteristics and the algorithm, the reader should refer to [14].

### 3.2 Comparison results for the current and the proposed outage modeling approach

Finally, in this section, the optimization results are presented for the two probabilistic modeling approaches presented in section 2.3. The purpose is to showcase that the optimal policies are significantly affected by the outage modeling used. In order to explore and examine the results, specific scenarios need to be defined. These scenarios correspond to the price movements in the DTMCs that are used to define the external feature of the state space. By using fixed price trajectories for both approaches, it is possible to compare them in an unbiased way: that means, the difference in the optimal policies is made exclusively by the different outage models. The scenario that is going to be examined here refers to a 420→310→167→150 $/kWh price trajectory for the Li-ion storage unit, 142→115→77→65 $/kWh for the lead-acid battery, 385→255→120→95 $/kWh for the vanadium redox and 3100→2600→1950→1700 $/kWh for the flywheel energy storage system.

Before presenting the results in Table 2 it is necessary to explain that an instance of the state, in accordance with the definition in Equation 1, is actually a tuple of 9 elements: the decision period, the four prices of the storage units and the four (already installed in the system) storage capacities. Table 2 is presented below:

Table 2: Optimal policies for both outage models

| Outage model / Decision period | | Single Poisson Process | Superposed Poisson Process |
|---|---|---|---|
| Period 1 | State | (0,420,142,385,3100,0,0,0,0) | (0,420,142,385,3100,0,0,0,0) |
| | Action | Do nothing | Do nothing |
| Period 2 | State | (1,310,115,255,2600,0,0,0,0) | (1,310,115,255,2600,0,0,0,0) |
| | Action | Add li-ion at 1000 kWh | Do nothing |
| Period 3 | State | (2,167,77,120,1950,1000,0,0,0) | (2,167,77,120,1950,0,0,0,0) |
| | Action | Add li-ion at 3000 kWh | Add li-ion at 1000 kWh |
| Period 4 | State | (3,150,65,95,1700,4000,0,0,0) | (3,150,65,95,1700,1000,0,0,0) |
| | Action | Add vanadium redox at 3000 kWh | Add li-ion at 3000 kWh |

It is clear from observing Table 2 that the optimal policies obtained from the two outage models differ significantly. The first thing that it should be noticed is the fact that the total storage capacity installed in the microgrid in the Single Poisson Process scenario is much higher (7000 kWh) than the corresponding total storage capacity installed in the Superposed Poisson Process scenario (4000 kWh). This finding could be attributed to the fact that in the latter scenario, the vast majority of outages are not long-lasting and therefore can be handled with a moderate amount of installed storage capacity.

However, this is obviously not the only difference existing in the two optimal policies. In the Single Poisson Process scenario, it can be observed that the first storage investment happens one decision period earlier than in the Superposed Poisson Process scenario. It indicates that under this scenario, lost demand during outages is costly enough so it not efficient to just endure the outages and it is better to proceed with storage installation earlier. Finally, it should also be mentioned that in the Single Poisson Process scenario, vanadium redox battery is chosen in the last decision period, while this is not the case for the Superposed Poisson Process scenario.

## 4. Conclusions

An existing reinforcement learning-based approach has been used in order to derive optimal policies for storage expansion in a microgrid setting. The focus of this work is to highlight the importance of analytical and accurate outage modeling, specifically tailored for the application that is being studied. In order to do so, two outage models are considered: one using a Single Poisson Process and another one using a Superposed Poisson Process. The algorithm is run for both approaches, with the rest of the parameters unchanged. The findings verify the hypothesis and show that further research is needed towards the direction of proper outage modeling.